\documentclass[a4paper,11pt]{article}
\pdfoutput=1 

\usepackage{jcappub} 

\usepackage{float}

\def\beq#1{\begin{equation}\label{#1}}
\def\eeq{\end{equation}}
\def\beqa#1{\begin{eqnarray}\label{#1}}
\def\eeqa{\end{eqnarray}}

\def\mycomment#1{\relax}
\title{\boldmath Instability of gravitational baryogenesis with fermions}

\author[a,b]{E. Arbuzova,}
\author[a,c]{and A. Dolgov}

\affiliation[a]{Novosibirsk State University,  Novosibirsk, 630090, Russia}
\affiliation[b]{Department of Higher Mathematics, Dubna State University, Dubna, 141980, Russia}
\affiliation[c]{ITEP,  Bol. Cheremushkinsaya ul., 25, 117218 Moscow, Russia}

\emailAdd{arbuzova@uni-dubna.ru}
\emailAdd{dolgov@fe.infn.it}

\abstract{The derivative coupling of baryonic current to the curvature scalar in gravitational baryogenesis 
scenarios leads to higher order equations for gravitational field. It is shown that these equations 
are strongly unstable and destroy standard cosmology. This is a generalization of our earlier 
results obtained  for scalar baryons to realistic fermions.   
}

\begin{document}
\maketitle
\flushbottom

\section{Introduction \label{s-intro}}

During the last decade scenarios of gravitational baryogenesis (GBG)~\cite{GBG-1} 
gained considerable popularity~\cite{GBG-more}.  They present a modification of the good
old scenario of spontaneous baryogenesis (SBG)~\cite{spont-BG} 
with the substitution instead of the 
(pseudo)goldstone field $\theta$ the curvature scalar $R$ coupled to the non-conserved
baryon current:  
\begin{eqnarray}
{\cal L}_{GBG} = \frac{f}{m_0^2} (\partial_\mu  R ) J^\mu_B\, ,
\label{L-GBG}
\end{eqnarray}
where $m_0$ is a constant parameter with dimension of mass and $f$ is dimensionless
coupling constant which is introduced to allow for an arbitrary sign of the above expression.

GBG scenarios possess the same interesting and nice features of SBG, namely generation
of cosmological asymmetry in thermal equilibrium without necessity  of explicit
C or CP violation in particle physics. However, an introduction of the derivative of
the curvature scalar into the Lagrangian of the theory results in high order gravitational
equations which are strongly unstable.  The effects of this instability may drastically 
distort not only the usual cosmological history, but also the standard Newtonian gravitational dynamics. In our recent paper~\cite{ea-ad-gbg-1} we discovered such
instability for  scalar baryons and here we found similar effect for the
more common spin one-half baryons (quarks).

\section{Equations of GBG }

We start from the action in the form:
\begin{eqnarray}
A= \int d^4x \sqrt{-g} \left[ \frac{m_{Pl}^2}{16 \pi} \,R - {\cal L}_{m}\right]\, 
\label{A-gen}
\end{eqnarray}
with  
\begin{eqnarray} \nonumber
{\cal L}_{m} &= &\frac{i}{2} (\bar Q \gamma ^\mu \nabla _\mu Q -   \nabla _\mu \bar Q\, \gamma ^\mu Q) - m_Q\bar Q\,Q\\ 
&+& 
\frac{i}{2} (\bar L \gamma ^\mu \nabla _\mu L -   \nabla _\mu \bar L \gamma ^\mu L)
- m_L\bar L\,L  \label{L-matt}\\ 
&+& \frac{g}{m_X^2}\left[(\bar Q\,Q^c)(\bar Q L) + (\bar Q^cQ)(\bar L Q) \right]
+ \frac{f}{m_0^2} (\partial_{\mu} R) J^{\mu} + {\cal L}_{other} , 
\nonumber
\end{eqnarray}
where $Q$ is the quark (or quark-like) field with non-zero baryonic number, $L$ is
another fermionic field (lepton),
$\nabla_\mu  $ is the covariant derivative of Dirac fermion in the tetrad formalism (see e.g.
lectures~\cite{Shapiro}), 
$J^{\mu } = \bar Q \gamma ^{\mu} Q$ is the quark current with $\gamma ^{\mu}$ being the curved space gamma-matrices,  
${\cal L}_{other}$ describes all other forms of matter.
The four-fermion interaction between quarks and leptons is introduced to ensure the
necessary non-conservation of the baryon number with $m_X$ being a constant
parameter with dimension of mass and $g$ being a dimensionless coupling constant.
In grand unified theories $m_X$ is usually of the order of $10^{14}-10^{15}$ GeV.

The Lagrangian (\ref{L-matt}) leads to the following equations of motion for quarks:
\begin{eqnarray} \nonumber
i \gamma ^\mu \nabla _\mu Q = m_Q Q -  \frac{f}{m_0^2} (\partial_{\mu} R) \gamma^{\mu} Q - 
\frac{g}{m_X^2}\left[2Q^c(\bar Q L) + (\bar Q Q^c)L \right]\,,\\
i \nabla _\mu \bar Q\, \gamma ^\mu = - m_Q \bar Q + \frac{f}{m_0^2} (\partial_{\mu} R) \bar Q \gamma^{\mu} +
\frac{g}{m_X^2}\left[2\bar Q^c(\bar L Q) + \bar L (\bar Q^c Q) \right]\, ,
\label{EoM-Q}
\end{eqnarray}
and leptons:
\begin{eqnarray} \nonumber
i \gamma ^\mu \nabla _\mu L = m_L L - \frac{g}{m_X^2}(\bar Q^c Q) Q\,,\\
i \nabla _\mu \bar L\, \gamma ^\mu = - m_L {\bar L}  +  \frac{g}{m_X^2}(\bar Q Q^c) \bar Q\,.
\label{EoM-L}
\end{eqnarray}
Note, that the fermionic part of Lagrangian (\ref{L-matt}), taken at the equations of
motion of quarks and leptons,
(\ref{EoM-Q}) and (\ref{EoM-L}), does not vanish due to the interaction between them:
\begin{eqnarray}
{\cal L}_{m} [\rm{Eqs \ of \ motion}] = - \frac{g}{m_X^2}\left[(\bar Q\,Q^c)(\bar Q L) + (\bar Q^cQ)(\bar L Q) \right]\,,
\label{lagr-EoM}
\end{eqnarray}
in contrast to the case of free fermions.

Taking variation of the action (\ref{A-gen}) over metric, $\delta A/ \delta g^{\mu \nu}$,  
we obtain the equations for gravitational field in the form:
\begin{eqnarray} 
\frac{m_{Pl}^2}{8\pi } \left( R_{\mu\nu} - \frac{1}{2} g_{\mu\nu} R \right) = T^m_{\mu \nu}\,,
\label{grav-def}
\end{eqnarray}
where energy-momentum tensor, $T^m_{\mu \nu}$, is defined as:
\begin{eqnarray}
 T^m_{\mu \nu} = \frac{ 2}{\sqrt{- g}} \frac{\delta A_m}{\delta g^{\mu \nu}}
 \label{T-def}
 \end{eqnarray} 
with
\begin{eqnarray}
A_m= \int d^4x \sqrt{-g} {\cal L}_{m}\, .
\label{A-gen}
\end{eqnarray}
The gravitational equations of motion, obtained this way, can be written as:
\begin{eqnarray} \nonumber
&&\frac{m_{Pl}^2}{8\pi }\left( R_{\mu\nu} - \frac{1}{2} g_{\mu\nu} R \right) = - g_{\mu \nu} {\cal L}_{m} +\\ \nonumber
&&
\frac{i}{4} \left[(\bar Q (\gamma _\mu \nabla _\nu + \gamma _\nu \nabla _\mu ) Q
 -    (\nabla _\nu \bar Q\, \gamma _\mu  + \nabla _\mu \bar Q\, \gamma _\nu )Q  \right] +\\ \nonumber
&& \frac{i}{4} \left[(\bar L (\gamma _\mu \nabla _\nu + \gamma _\nu \nabla _\mu ) L
 -    (\nabla _\nu \bar L\, \gamma _\mu  + \nabla _\mu \bar L\, \gamma _\nu )L  \right] -\\ 
 && \frac{2f}{m_0^2} \left[R_{\mu\nu} + g_{\mu \nu } D^2 - D_{\mu} D_{\nu}\right] D_{\alpha} J^{\alpha} + 
 \frac{f}{2m_0^2} (J_{\mu} \partial_{\nu}R +  J_{\nu} \partial_{\mu}R)\,,
 \label{EoM-grav}
 \end{eqnarray}
 where $D_\mu$ is the usual tensor covariant derivative in the background metric.
 
Taking trace of equation (\ref{EoM-grav}) with respect to $\mu$ and $\nu$ we obtain:
\begin{eqnarray} \nonumber
- \frac{m_{Pl}^2}{8\pi } R &=& m_{Q} \bar Q Q + m_L \bar L L 
+
 \frac{2g}{m_X^2}\left[(\bar Q\,Q^c)(\bar Q L) + (\bar Q^cQ)(\bar L Q) \right] \\
 &-& 
 \frac{2f}{m_0^2} (R + 3D^2) D_{\alpha} J^{\alpha} + T_{other}\,,
 \label{trace}
 \end{eqnarray}
where $T_{other} $ is the trace of the energy momentum tensor of all other fields.
At relativistic stage, when masses are negligible, we can take  $T_{other} = 0$. 
The average  expectation value of the interaction term in eq. (\ref{trace}), which is 
proportional to $g$, is also 
small, especially at $T<m_X$, so the contribution of all matter fields may be neglected.

We see in what follows, that the kinetic equation (\ref{kin-eq-gnrl})
leads to an explicit dependence 
of the current divergence, $D_\alpha J^\alpha$, on $R$, if the current is not conserved. As
a result we obtain high (fourth)  order equation for $R$, as is discussed in the 
next section.

We can use an alternative  representation of the quark field:
\begin{eqnarray} 
Q_2 = \exp( i f R /m_0^2 )\, Q
\label{Q2}
\end{eqnarray}
analogously to what is done in our paper \cite{ADN}.
The substitution of $Q_2$ instead of $Q$ results in the  elimination of the term $ f J^{\mu} \partial_{\mu}R /m_0^2 $ in the
Lagrangian (\ref{L-matt}) 
but the dependence on the curvature reappears in the interaction term as:
\begin{eqnarray}
 \frac{2g}{m_X^2}\left[ e^{-3ifR/m_0^2}\, (\bar Q_2\,Q_2^c)(\bar Q_2 L) + 
 e^{3ifR/m_0^2}\,(\bar Q_2^cQ_2)(\bar L Q_2) \right] .
\label{int-R}
\end{eqnarray}
Nevertheless we obtain the same fourth order equation for the evolution of curvature,
as in the case of the non-rotated field $Q$.

In what follows we study solutions of eq. (\ref{trace}) in cosmology in homogeneous and
isotropic FRW background with the metric: 
\begin{eqnarray}
ds^2=dt^2 - a^2(t) d{\bf r}^2\, .
\label{ds-2}
\end{eqnarray}
In this background the curvature is a function of time only and the covariant derivative acting on a vector
$V^\alpha (t) $, which has only time component, has the form:
\begin{eqnarray}
D_\alpha V^\alpha = (\partial_t + 3 H) V^t ,
\label{DV}
\end{eqnarray}
where $H = \dot a/a$ is the Hubble parameter.

\section{Kinetic equation \label{kin-eq}}

Lat us consider e.g.  the reaction $q_1 + q_2 \leftrightarrow \bar q_3 + l_4$, where $q_1$ and
$q_2$  are quarks with momenta $q_1$ and $q_2$, while $\bar q_3$ and $l_4$ are antiquark
and lepton with momenta $q_3$ and $l_4$. We use the same notations for the particle
symbol and for the particle momentum.
The kinetic equation for the variation of the baryonic number density $n_B \equiv J^t$
through this reaction in the FRW background has the form: 
\begin{eqnarray}
(\partial_t +3 H) n_B = I_B^{coll}, 
\label{kin-eq-gnrl}
\end{eqnarray}
where the collision integral for space and time independent interaction is equal to:
\begin{eqnarray} 
&&I^{coll}_B =- 3 B_q (2\pi)^4  \int  \,d\nu_{q_1,q_2}  \,d\nu_{\bar q_3, l_4}
\delta^4 (q_1 +q_2 -q_3 - l_4)
\nonumber\\
&& \left[ |A( q_1+q_2\rightarrow  \bar q_3 +l_4)|^2
f_{q_1} f_{q_2} -
 |A( \bar q_3 +l_4  \rightarrow  q_1+q_2 ) |^2
 f_{\bar q_3} f_{l_4}
 \right],
\label{I-coll}
\end{eqnarray}
where $ A( a \rightarrow b)$ is the amplitude of the transition from state $a$ to state $b$,
$B_q$ is the baryonic number of quark, $f_a$ is the phase space distribution (the
occupation number), and 
\begin{eqnarray}
d\nu_{q_1,q_2}  =  
\frac{d^3 q_1}{2E_{q_1} (2\pi )^3 }\,  \frac{d^3 q_2}{2E_{q_2} (2\pi )^3 } ,
\label{dnuy}
\end{eqnarray}
where $E_q = \sqrt{ q^2 + m^2}$ is the energy of particle with three-momentum
$q$ and mass $m$. The element of phase space of final particles, $d\nu_{\bar q_3, l_4} $, is defined analogously. 

We neglect the Fermi suppression factors and the effects of gravity
 in the collision integral. This is generally a good approximation.

The calculations are strongly simplified if quarks and leptons are in equilibrium
with respect to elastic scattering and annihilation. In this case their distribution functions 
take the form
\begin{eqnarray}
f = \frac{1}{e^{(E/T - \xi}) + 1} \approx e^{-E/T + \xi}, 
\label{f-eq}
\end{eqnarray}
where $\xi = \mu/T$ is dimensionless chemical potential, different for quarks, $\xi_q$,
and leptons, $\xi_l$. 

The assumption of kinetic equilibrium is well justified since it is usually enforced by 
very efficient elastic scattering. Equilibrium with respect to annihilation, say, 
into two channels: $2\gamma$ and $3\gamma$, implies 
the usual relation between chemical potentials of particles and antiparticles,
$\bar \mu = -\mu$. However, if we use the original representation for the quark
fields, when they satisfy equations of motion (\ref{EoM-Q}), the conclusion of 
kinetic equilibrium is not evident because the quark evolution depends upon 
$R (t)$, which may be quickly varying, as we see in what follows.  At first sight
the equilibrium distribution may not be able to keep pace with a fast variation of $R$.
This problem is absent in the representation (\ref{Q2}), since $R(t)$  neither
enters the equation of motion, nor the amplitudes of elastic scattering and annihilation.

In representation (\ref{Q2}) the baryonic number density is given by the expression:
\begin{eqnarray} \nonumber
n_B &=& \int \frac{d^3 q}{2 E_q\, (2\pi)^3}  (f_q - f_{\bar q})  \\
&=& \frac{g_S B_q}{6} \left(\mu T^2 + \frac{\mu^3}{ \pi^2}\right) =
\frac{g_S B_q T^3}{6}\,\left(\xi + \frac{\xi^3}{\pi^2}\right) \,,
\label{n-B-of-mu}
\end{eqnarray}
where  $T$ is the cosmological plasma temperature, $g_S$ and $B_q$ are respectively
 the number of the spin states and the baryonic number of quarks. 
 
Since the transition amplitudes, which enter the collision integral, are obtained by
integration over time of the Lagrangian operator (\ref{int-R}), taken
between the initial and final states, the energy conservation delta-function in 
eq. (\ref{I-coll}) would be modified due to time dependent factors $\exp [\pm 3 ifR(t)/m_0^2]$. 
In the simplest case, which is usually considered
in gravitational (and spontaneous) baryogenesis, a slowly
changing $\dot R$ is taken, so we can approximate $R(t) \approx \dot R(t)\,t$.  
For a constant $\dot R$ the energy is not conserved but the energy
conservation condition is trivially modified, as 
\begin{eqnarray} \nonumber
&&\delta [ E(q_1) + E(q_2) - E(q_3) - E(l_4) ] \rightarrow \\
&&\rightarrow \delta [ E(q_1) + E(q_2) - E(q_3) - E(l_4) - 3f\dot R(t) /m_0^2\,]. 
\label{E-non-conserved}
\end{eqnarray}
Thus the energy is non-conserved due to the action of the external field $R(t)$.
Delta-function (\ref{E-non-conserved}) is not precise, but the result is pretty close to it,  
 if $\dot R(t)$    changes very little
during the effective time of the relevant reactions. 

If the dimensionless chemical potentials $\xi_q$ and $\xi_l$, as well as 
$ f\dot R(t) /m_0^2 /T $, are small, and 
 the energy balance is ensured by the delta-function (\ref{E-non-conserved}),
the collision integral can be approximated as:
\begin{eqnarray}
I^{coll}_B \approx \frac{C_I g^2 T^8}{m_X^4}\,  
\left[ \frac{3 f\dot R(t)}{ m_0^2\,T} - 3\xi_q + \xi_l  \right] , 
\label{Icoll-appr}
\end{eqnarray}
where $C_I$ is a positive dimensionless constant. 
The factor $T^8$ appears for reactions with massless particles and 
the power eight is found from dimensional consideration. 
Because of conservation of
the sum of baryonic and leptonic numbers $\xi_l = -\xi_q/3 $.

The case of an essential variation of $\dot R (t)$ is analogous to fast
variation of $\dot \theta (t)$ studied in our paper~\cite{ADN}. 
Clearly, it is  much more complicated technically. Here we consider only the simple 
situation with quasi-stationary background and postpone more realistic time
dependence of $R(t)$ for the future work.

For small chemical potential the baryonic number density (\ref{n-B-of-mu}) is equal to 
\begin{eqnarray}
n_B \approx \frac{g_s B_q}{6}\, \xi_q T^3\, ,
\label{n-small-xi}
\end{eqnarray}
and if the temperature adiabatically decreases
in the course of the cosmological expansion, according to $\dot T = - H T$, equation
(\ref{kin-eq-gnrl}) turns into
\begin{eqnarray}
\dot \xi_q = \Gamma \left[ \frac{9 f\dot R(t)} {10m_0^2\, T} - \xi_q \right],
\label{dot-xi-q}
\end{eqnarray}
where $\Gamma \sim g^2 T^5/m_X^4 $ is the rate of B-nonconserving reactions.

If $\Gamma$ is in a certain sense large, this equation can be solved in stationary
point approximation as
\begin{eqnarray} 
\xi_q=\xi_q^{eq} - \dot \xi_q^{eq}/\Gamma\, , 
\label{xi-approx}
\end{eqnarray}
where
\begin{eqnarray}
\xi_q^{eq} = \frac{9}{10} \frac{f \dot R} {m_0^2 T}\, .
\label{xi-q-equil}
\end{eqnarray}  
This is the main conclusion of this section. 
If we substitute $\xi_q^{eq} $ (\ref{xi-q-equil}) into eq.~(\ref{trace}) we arrive 
to the fourth order equation for $R$, as it is described in the next section.

\section{Curvature instability \label{sec-R}} 

According to the comment below eq. (\ref{trace}), the contribution of thermal matter
into this equation can be neglected and we arrive to the very simple fourth order
differential equation:
\begin{eqnarray}
\frac{d^4 R}{dt^4} = \lambda^4 R,
\label{d4-R}
\end{eqnarray}
where $\lambda^4 = C_\lambda m_{Pl}^2 m_0^4 /T^2 $ with
$ C_\lambda  = 5/(36\pi f^2 g_s B_q )$. Deriving this equation we neglected
the Hubble parameter factor in comparison with time derivatives of $R$. It is justified a posteriori because
the calculated $\lambda$ is much larger than $ H $.

Equation (\ref{d4-R}) has the exponential solutions $R \sim \exp (\mu_n t ) $ with 
\begin{equation}
\mu_n = \lambda \,\exp (i n \pi/ 2)  
\label{mu}
\end{equation}
with $ n= 0,1,2,3$.
Evidently this equation 
has extremely unstable solution with instability time
by far shorter than the cosmological time. This instability would lead to an explosive
rise of $R$, which may possibly be terminated  by the nonlinear  terms proportional to the
product of $H$ to lower derivatives of $R$. Correspondingly one may expect stabilization when $HR \sim \dot R$,
i.e. $H\sim \lambda$. Since 
\begin{eqnarray}
\dot H + 2 H^2 = - R/6,
\label{dot-H}
\end{eqnarray}
$H$ would also exponentially  rise together with  $R$, 
$ H \sim \exp (\lambda t )$ and $\lambda H \sim R$.
Thus stabilization may take place at $R \sim \lambda^2   \sim  m_{Pl} m_0^2 / T$.
This result is much larger than the normal General Relativity value
$ R_{GR} \sim T_{matter} /m_{Pl}^2 $, where $T_{matter}$ is the trace of the 
energy-momentum tensor of matter.

If $\dot R$ is still slow, such that the energy balance condition (\ref{E-non-conserved}) is fulfilled,
but  $\dot R/(m_0^2 T)$ is large, then the asymmetry 
would also become large and the approximation of Boltzmann statistics becomes
invalid. Nevertheless the equilibrium solution which annihilates the 
collision integral remains the same,  (\ref{xi-q-equil}),
but for $\xi \gtrsim 1$ the cubic terms in the baryonic density (\ref{n-B-of-mu}) becomes
essential and instead of equation (\ref{d4-R}) we obtain
\begin{eqnarray}
\frac{d^2}{dt^2} \left[ \ddot R \left( 1+ \frac{3}{\pi^2}
\left(\frac{9 f \dot R}{10 m_0^2 T} \right)^2 \right) \right]  = \lambda^4 R\,.
\label{d2-R2}
\end{eqnarray}

If the lower order derivatives of $R$ still can be neglected we arrive to the simpler equation
\begin{eqnarray}
\frac{d^4 R}{dt^4} = \lambda^4 R\,\left[ 1 + \frac{243}{10\pi^2} 
\left(\frac{f \dot R}{m_0^2 T}\right)^2  \right]^{-1},
\label{d4-R2}
\end{eqnarray}
from which it is evident that the rise of $R$ should terminate when 
$ \dot R \sim m_0^2 T/f $.

\section{Discussion and conclusion \label{sec-concl}}

The considered here effect of strong instability in high order differential equations with small
coefficient $\epsilon$ in front of the highest derivative term is well known in mathematics but might
be unexpected for a physicist. Even more surprising than the instability 
is a discontinuity of the limit $\epsilon \rightarrow 0$. If we take
$\epsilon = 0$ from the very beginning, then the instability does not appear and the theory is
reduced to the normal lower order one, while with any small but non-zero $\epsilon$ the equation of
motion has solutions which are absent for $\epsilon = 0$. Moreover, the smaller is $\epsilon$, the faster
is the rise of the unstable solution. Surprising at first sight, this is very well established fact,
as one can check playing with simple model examples of higher order differential equations.

There is an apparent counterexample known in quantum field theory, namely, the decoupling of heavy modes.
The low energy limit of a normal field theory is not sensitive to existence of very
high mass particles. However, this is true only for stable equations of motion. The equations of motion may be higher order
but the key words is "stable".  In the case considered in the present work the condition of stability is violated.
The instability may be present at any $\epsilon$, even  very large one, but, we repeat, that the rate of the instability 
development is faster at smaller $\epsilon$.


The found here effect of strong instability of high order differential equations with small coefficient, $\epsilon$,
in front of the highest derivative term, follows from the well known Lyapunov stability theory. In its classical form this theory 
is applied to the solutions of a system of generally non-linear first order differential equations. 
The infinitesimal variation of the solution under scrutiny leads to a homogeneous system of linear differential equations
satisfied by these small variations.
The properties of the solutions are determined by the eigenvalues of the characteristic determinant of the emerging 
system of these linear equations.  If the eigenvalues are all negative or imaginary, then the solution is stable. Small fluctuations 
around it do not rise with time. Real positive eigenvalues induce an exponential rise of small fluctuations with time, leading to  
a strong deviation from the original solution. In this work we studied the issue of stability of the gravitational equations of motion 
which are reduced to the linearized fourth order equation (\ref{d4-R}). This equation 
can be trivially transformed to the classical Lyapunov system, though it is not necessary. All Lyapunov eigenvalues can be
determined directly from eq. (\ref{d4-R}). One of the  eigenvalues $\mu_n$ (\ref{mu})
is positive and huge, so the rise of $R$ is really fast. 

For the subsequent discussion is convenient  to consider the toy model governed by the linear 4th order equation of the form:
\begin{eqnarray}
\epsilon  \,\frac{ d^4 R}{dt^4}  + c R = 0.
\label{d4-R-2}
\end{eqnarray}
The case $\epsilon = 0$ in this toy example corresponds to the General Relativity limit.

Equations of similar kind arise  in the models of F(R) modified gravity, suggested for the 
description of the observed cosmological acceleration. The  effects established 
in these works are similar to those found in the present paper. 
Usually the coefficients in front of the highest derivative in such equations are assumed to be small. By assumption,
in our toy model the approach to GR
is expected, when $\epsilon \rightarrow 0$.   If the equation of motion of modified theory is stable with respect to small 
perturbations, the limiting transition to GR is realized without problems. However, if the equation is unstable there is no transition to GR 
when $ \epsilon $ tends to zero, remaining small but  non-vanishing. So the theories with $\epsilon = 0$ and with $\epsilon$ 
arbitrary close to zero are  very much different. There  is no  continuous limit  as $\epsilon \rightarrow 0$.

Usually the modified gravitational equations are of higher order with respect to unmodified GR ones. 
If the former are unstable,
the phenomenological implications of modified theories may be endangered. The applicability of such theories is determined by the
characteristic time of the instability,  $ t_{in} $.  If $ t_{in}$ is sufficiently long, one may not worry about this instability, since it evolves 
very slowly to be observed. So we are in some kind of the crunch: large $\epsilon$ means large deviation from GR, while small $\epsilon$ may 
lead to fast instability if the equation is unstable. Such phenomenon was observed in $ F(R) \sim 1/R^n $~\cite{1-over-R} 
theories as found in the work~\cite{AD-MK}. 
 Accordingly to cure this type of instability the function $F$ was modified in the works~\cite{AAS-F-of-R}. 
 However, further modification was found to be necessary because of emergence of past~\cite{ABS} 
 and future singularities~\cite{AVF}.  

The kind of instability found in our work on gravitational baryogenesis is 
mathematically the same as that  discovered in the examples presented above.
In all the unstable cases,  studied in the published works,
the limiting transition to GR with vanishingly small modification was not possible because of exponentially fast instability. The instability 
may be terminated due to non-linear terms in the equation but usually the stabilization takes place at the values of R much larger than 
the canonical GR values. Though initially the corrections to GR may be very small, proportional to the small coefficient $\epsilon$, they rise
exponentially and soon destroy GR. In this sense there is no limiting transition to GR at small $\epsilon$.

Returning to the 4th order equation (\ref{d4-R-2}), we  see that the eigenvalues are $ \mu_n = (-c/ \epsilon)^ {1/4} $. This equation has 4 roots, n=0,1,2,3,
see eq. (\ref{mu}),
and at least one of them has positive real part, independently on the sign of $ c/\epsilon $. There is a known pathology in 
$\lambda \phi^4 $ - theory with negative $\lambda$. But in this example the equation of motion is second order and so 
is not directly related  to our case.

There is another example of instability of high order equations, the so called Ostrogradsky instability. His original work is difficult to find but there is a good review of this instability in the paper~\cite{wood}. 
In particular, an example of the Lagrange theory with higher derivative is considered there. The corresponding fourth order equation of 
motion (29) of the paper~\cite{wood} is 
stable in the Lyapunov sense due to a special relation between the coefficients of the equation. 
In our case there is no such a relation and judging by 
the value of $\mu_n$ our equation is strongly unstable, in the sense that
the instability is developed in very short time. Possibly the solution is stabilized by the nonlinearity of the equation, but at very high curvature.

So the simple version of the gravitational baryogenesis does not work, because of the found here instability. 
However, we cannot exclude that some further modification of this scenario may 
possibly cure its sickness. 

In this work we have described only the basic features of the new effect of instability in gravitational baryogenesis
with fermions. For a more accurate analysis numerical solution will be necessary. We plan to perform it in another work. 
The problem is very complicated technically, because the assumption of slow variation of $\dot R$ quickly becomes broken 
and the collision integral should be evaluated in time dependent background. 
This is by far
not so easily tractable as the usual stationary one.  We will also take into account finite integration limits over time.
The technique for solving kinetic equation in non-stationary
background is developed  in ref.~\cite{ADN}. 

To conclude we have shown that gravitational baryogenesis in the simplest versions
discussed in the literature is not realistic because the instability
of the emerging gravitational  equations destroys the standard cosmology. Some
stabilization mechanism is strongly desirable. Probably stabilization may be achieved in a version
of $F(R)$ modified gravity or by an introduction the formfactor $g(R)$  into the coupling (\ref{L-GBG}), such that
$g(R)$ drops down with rising $R$.\\[4mm]

\section*{Acknowledgements}
We acknowledge the support of the Grant of President of Russian Federation for the leading scientific Schools of Russian Federation, NSh-9022-2016.2.

\end{document}